\documentclass[draftclsnofoot,onecolumn]{IEEEtran}

\usepackage{amsthm,amsmath,amssymb,cite}
\usepackage{eucal}
\usepackage{xifthen}
\usepackage{mathtools}
\usepackage{enumerate}
\usepackage{microtype}
\usepackage{xspace}
\usepackage{bm}
\usepackage{lastpage}
\usepackage[center]{caption}
\usepackage{xcolor}
\allowdisplaybreaks
\usepackage{enumitem}

\newcommand{\uu}{\mbox{UE}}
\newcommand{\en}{\mbox{EN}}

\newtheorem{definition}{Definition}

\newtheorem{lemma}{Lemma}

\newtheorem{example}{Example}
\newtheorem{proposition}{Proposition}
\newtheorem{corollary}{Corollary}

\DeclareMathAlphabet{\mathcal}{OMS}{cmsy}{m}{n}

\usepackage{graphicx}
\usepackage[font=small]{caption}
\usepackage{subcaption}

\begin{document}

\title{Cache-Aided Content Delivery in Fog-RAN Systems with Topological Information and no CSI}


\author{
Wei-Ting~Chang, Ravi~Tandon, Osvaldo~Simeone
\thanks{W.-T.~Chang and R. Tandon are with the Department of Electrical and Computer Engineering at the University of Arizona, Tucson, AZ, USA (email: \{wchang, tandonr\}@email.arizona.edu).}
\thanks{O.~Simeone is with the Department of Informatics at King's College London, London, UK (email: osvaldo.simeone@kcl.ac.uk).}
\thanks{O.~Simeone has received funding from the European Research Council (ERC) under the European Union's Horizon 2020 Research and Innovation Programme (Grant Agreement No. 725731). His work was also partially supported by the U.S. NSF through grant 1525629.}
\thanks{This work was presented at the 2017 51st Asilomar Conference on Signals, Systems and Computers, Pacific Grove, California, USA.}}

\maketitle

\begin{abstract}
In this work, we consider a Fog Radio Access Network (F-RAN) system with a partially connected wireless topology and no channel state information available at the cloud and Edge Nodes (ENs). An F-RAN consists of ENs that are equipped with caches and connected to the cloud through fronthaul links. We first focus on the case where cloud connectivity is disabled, and hence the ENs have to satisfy the user demands based only on their cache contents. For a class of partially connected regular networks, we present a delivery scheme which combines intra-file MDS coded caching at the ENs and blind interference avoidance on the wireless edge channel. This scheme requires minimum storage and leads to an achievable Normalized Delivery Time (NDT) that is within a constant multiplicative factor of the best known NDT with full storage. We next provide achievable schemes for the case where cloud connectivity is enabled, and we provide new insights on the interplay between cloud connectivity and edge caching when only topological information is available.

\end{abstract}


\section{Introduction}
The growth in the demand of multimedia contents is increasing the need to reduce content delivery latency in wireless networks, which is one of the goals of 5G \cite{5GSurvey}. To tackle this challenge, edge caching is emerging as one of the main candidate solutions. Edge caching enables the Edge Nodes (ENs) in a wireless system, such as base stations or access points, to locally store popular content. By prefetching popular content during off-peak hours or online, the ENs can deliver these content without retrieving them from core networks, thus reducing latency.

Caching for wireless networks is an active area of recent research. Cache-aided interference channels with full Channel State Information (CSI) in the case of three ENs and three users were studied in \cite{cacheIC}. A performance upper bound in terms of the inverse of sum Degrees of Freedom (DoF) for this setting was obtained in \cite{cacheIC}. Information-theoretic lower bounds for cache-aided wireless networks with full CSI for any number of ENs and users were found in \cite{SenTan2016}. Edge caching was investigated in the presence of caches at the receivers in \cite{HacNieDig2016,XuTaoLiu2016,NadMadAve2017, RoiTosGun2017}. A novel approach that achieves approximately optimal DoF by separating physical and network layers was presented in \cite{HacNieDig2016}. Upper and lower bounds on the minimum Normalized Delivery Time (NDT), which is related to the inverse of the DoF, were provided in \cite{XuTaoLiu2016}, and their optimality was shown for certain cache storage regimes. The scheme of \cite{NadMadAve2017} was shown to be within a constant multiplicative factor of $2$ of the optimal in the presence of full CSI under linear precoding.

Edge caching was then studied in combination with cloud-aided transmission to yield the Fog Radio Access Network (F-RAN) architecture in \cite{TanSim2016}. There, the optimal NDT trade-off was characterized for two ENs and two users. More general upper and lower bounds on the NDT were derived in \cite{SenTanSim2017} for any number of ENs and users, characterizing the minimum NDT within a multiplicative factor of $2$. Online caching in F-RANs was studied in \cite{AziSimSen2017} and scenario with heterogeneous contents was considered in \cite{GosSimPop2017}.

The prior works summarized above on F-RAN assume that the wireless network is fully connected and that all the ENs and the cloud have full CSI. The assumption of full connectivity and CSI may not be valid in practice. In fact, links between ENs and users may be too weak due to large geographic separation or severe fading. As an approximation, nodes with such weak links can be viewed as disconnected from each other. Under such an approximation, one obtains a partially connected topology. Furthermore, full CSI requires significant overhead, particularly for large networks. To alleviate these assumptions, we study the extreme case in which cloud and ENs have no CSI but only knowledge of wireless network topology.

Interference management techniques for partially connected network with only topology knowledge have been studied under the rubric of Topological Interference Management (TIM), TIM has interesting connection to index coding \cite{JafarIndex2014}. Cooperative TIM (or TIM-CoMP) was studied in \cite{YiGes2015}, by allowing ENs to cooperate with each other. Achievable schemes and upper bounds on the NDT under TIM-CoMP setting were provided in \cite{YiGes2015} (also see \cite{LamZhaEli2017} for another variation of cooperative TIM). The work of \cite{YiCai2016} is the closest to the problem consider in this paper. In \cite{YiCai2016}, the ENs can only store a subset of files in the library. This reference focused on blind zero-forcing based interference management and studies the minimization of the delivery latency from an algorithmic standpoint.

\textit{Main Contributions:} In this paper, we study a class of partially connected networks, referred to as $(K,d)$ regular network, with no CSI at the cloud and ENs. In this network, as seen in Fig. \ref{model}, there are $K$ ENs and $K$ users, and each EN is connected to the corresponding user and the subsequent $d-1$ users in a cyclic fashion in the wireless channel. During the offline pre-fetching phase, the ENs can cache a function of the files in the library of popular contents subject to storage constraints. In addition, the ENs are connected to the cloud with finite capacity fronthaul links. Users can request any file in the library, and the requested files must be recovered through the transmission over the wireless channel.

This paper addresses the following questions: What is the minimum necessary cache storage at each EN so that the users can reliably decode their desired files when there is no cloud connectivity? What caching strategies allow the ENs to leverage the topological knowledge of a network? How to devise delivery schemes under cache storage constraint at each EN and fronthaul capacity constraint? The specific main contributions are as follows.

\begin{itemize}
\item We first study the case with $\mu=1/d$ and no cloud connectivity. We propose a Maximum Distance Separable (MDS) coded caching scheme that ensures that each user can recover any arbitrary requested file by means of a blind interference avoidance-based delivery scheme. We characterize the corresponding achievable NDT as a function of connectivity degree $d$ and cache storage size $\mu$. We show that the resulting NDT is within a constant multiplicative factor of $4$ of the best known NDT with full caching proposed in \cite{YiGes2015}.
\item For the case when cloud connectivity is enabled, we present different caching and delivery strategies as a function of cache sizes and fronthaul capacity. In the low and medium cache size regimes, if the fronthaul capacity is low, the proposed scheme is shown to obtain a lower NDT than the scheme in \cite{YiGes2015} by means of cloud transmission despite the fronthaul overhead.
\end{itemize}

The remainder of the paper is organized as follows. We describe the $K$-user F-RAN model under study in Section \ref{sys}. In Section \ref{main}, we present the proposed scheme in the absence of cloud connectivity. Section \ref{discussion} discusses the full F-RAN model where cloud processing is enabled. We conclude the paper in Section \ref{conclusion}.

\section{System Model and Preliminaries \label{sys}}

\begin{figure}[h]
\centering
	\includegraphics[width=0.7\linewidth]{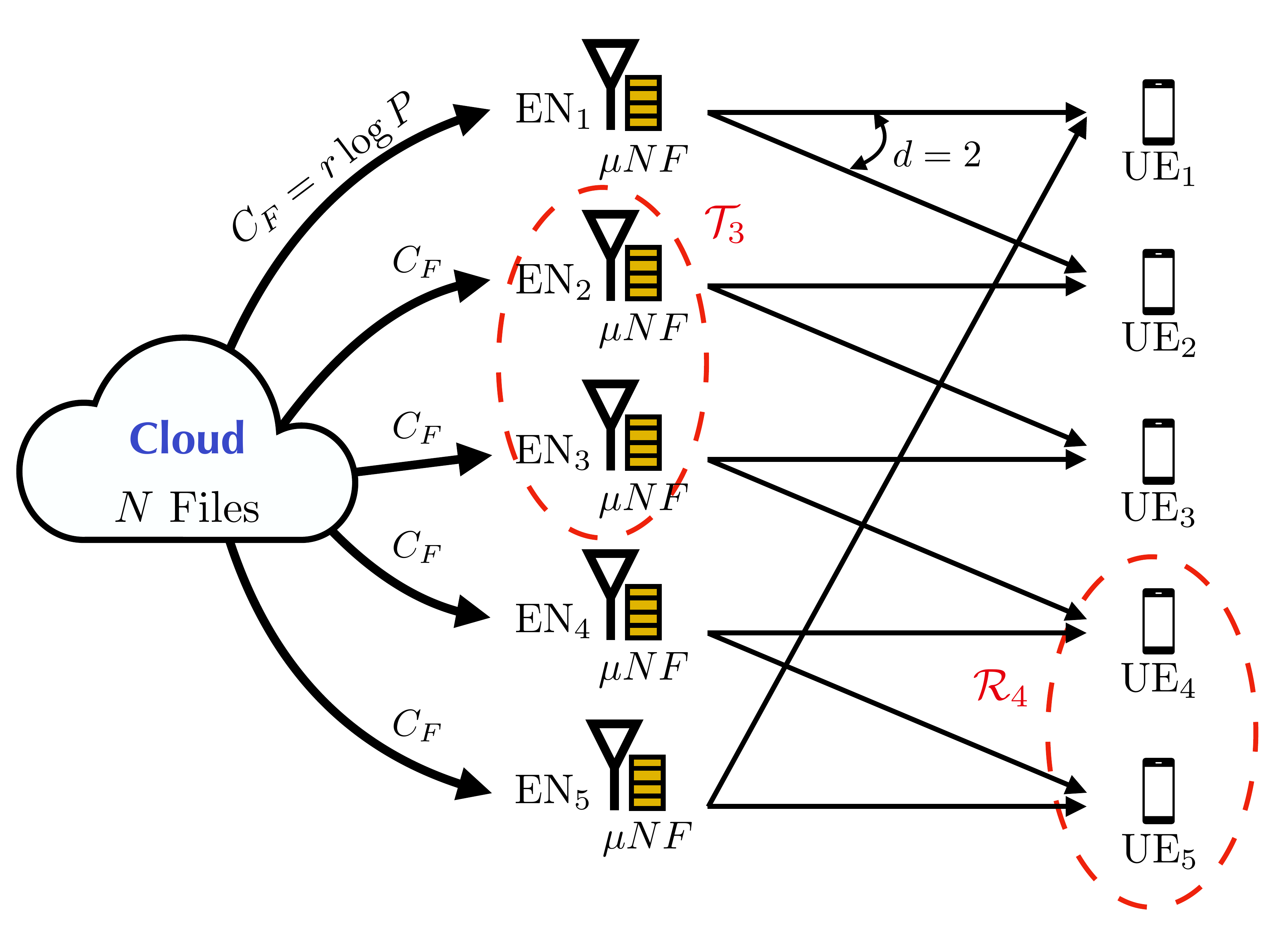}
	\vspace{-25pt}
	\caption{A $(5,2)$ regular F-RAN model.
	\label{model}}
	\vspace{-15pt}
\end{figure} 
We consider a partially connected $K$-user F-RAN network with one antenna at each EN and user and connectivity degree $d$, as illustrated in Fig. \ref{model}. $\mathcal{G}$ represents the network topology, where $\mathcal{G}$ is defined by the sets $(\mathcal{T}_1,\mathcal{T}_2,\dots,\mathcal{T}_K,\mathcal{R}_1,\mathcal{R}_2,\dots,\mathcal{R}_K)$. Specifically, the set of ENs that are connected to user $\uu_j$ is denoted as $\mathcal{T}_j$, and the set of users that are connected to $\mbox{EN}_i$ is denoted as $\mathcal{R}_i$. If each $\mbox{EN}_i$ transmits a symbol $X_i(t)$, the received signal at user $\uu_j$ at time instant $t$ is
\begin{equation}
Y_j(t)=\sum_{i\in\mathcal{T}_j}H_{j,i}(t)X_i(t)+U_j(t),\label{transmitSig}
\end{equation}
where $H_{j,i}(t)$ is the channel coefficient between $\mbox{EN}_i$ and $\uu_j$ at time instant $t$. The channel coefficients are assumed to be drawn from a continuous distribution, and are independent and identically distributed (i.i.d.) across all $t$ and $k$. The additive white Gaussian noise $U_j(t)$ is assumed to have zero mean and unit variance and independent to all variables $H_{j,i}(t)$ and $X_i(t)$. Each EN is connected to a cloud processor through a fronthaul link of capacity $C_F$ bits per symbol of the wireless channel.

Each EN has a local cache of size $\mu NF$ bits, where $\mu\in[0,1]$ is the fractional cache size. We denote $Z_k$ as the cache content stored at $\mbox{EN}_k$ during the pre-fetching phase, and we focus on cases in which only intra-file coding is allowed. Accordingly, the cached content after the pre-fetching phase is given as $Z_k=(Z_{k,1},Z_{k,2},\dots,Z_{k,N})$, where $Z_{k,n}$ is a function of the $n^{th}$ file $W_n$ and each $Z_{k,n}$ is of size
\begin{equation}
H(Z_{k,n})\leq\mu F, \forall n=1,\dots,N, k=1,\dots,K.
\end{equation}

At the start of the delivery phase, each user requests one file from the library, and the demand vector is denoted as $\overline{D}\triangleq(d_1,d_2,..,d_K)$. We note that the cache content does not depend on the user's demands. We focus on F-RANs that operate in serial fashion: once the demand vector is revealed, the cloud first sends the data through individual fronthaul links to the ENs, and then the ENs transmit signals through wireless channel (\ref{transmitSig}). The latency is then the sum of latency of fronthaul transmission and latency of wireless transmission. The block length of the channel code is denoted by $T_E$ and the received block signal at $\uu_j$ is
\begin{equation}
Y_j^{T_E}=\sum_{i\in\mathcal{T}_j}H_{j,i}^{T_E} X_i^{T_E}+U_j^{T_E},
\end{equation}
where we impose the average power constraint $T_E^{-1}E(|X_i^{T_E}|^2)\leq P$. Any feasible sequence of policies must satisfy the following worst-case constraint on the probability of error
\begin{equation}
\max\limits_{\overline{D}}\max\limits_{k} \text{Pr}(W_{d_k}\not=\widehat{W}_{d_k})\leq\epsilon,\label{error}
\end{equation}
for any $\epsilon>0$, i.e., where $\widehat{W}_{d_k}$ is the decoded file at $\uu_{k}$. Condition (\ref{error}) imposes that the probability of decoding error across all users and over all possible demand vectors $\overline{D}$ be made arbitrarily small as $T_E\rightarrow\infty$.

Throughout the paper, we focus on a class of partially connected regular edge channels as defined next.

\begin{definition} \label{Reg}(Regular Network)
In regular $(K, d)$ symmetric partially connected edge channel, the set of users that are connected to $\mbox{EN}_i$ is defined as $\mathcal{R}_i\triangleq\{i,i+1,\dots,i+(d-1)\}$, for all $i\in [K]$, and the set of ENs that are connected to $\uu_j$ is defined as $\mathcal{T}_j\triangleq\{j-d+1, j-d+2,\dots,j\}$, for all $j\in [K]$, where all indices are modulo $K$. To be specific, every $\mbox{EN}_i$ is connected to its corresponding user and $d-1$ subsequent users in cyclic manner, where $1\leq d\leq K$. 
\end{definition}

We note that, for a $(K,d)$ regular network without cloud connectivity, the fractional cache size $\mu$ must be at least $1/d$ for reliable decoding, i.e. $1/d \leq\mu\leq 1$. In fact, when $\mu<1/d$ and cloud connectivity is disabled, a user can receive at most $d\mu F$ bits of a file from its $d$ connected ENs. Therefore, since $d\mu F<F$, some bits of a file cannot be received to a user.

In this paper, we use NDT as the performance metric, which is defined as follows \cite{SenTanSim2017}.

\begin{definition}
(Delivery time per bit). For a given sequence of feasible policies, an achievable delivery time per bit $\Delta(\mu,C_F,d,P)$ is given as
\begin{equation}
\Delta(\mu,C_F,d,P)=\lim\limits_{F\rightarrow\infty}\frac{T_F+T_E}{F}, \label{DTB}
\end{equation}
with fronthaul and edge contributions given as $\Delta_E(\mu,C_F,d,P)=\lim\limits_{F\rightarrow\infty}\frac{T_E}{F}$ and $\Delta_F(\mu,C_F,d,P)=\lim\limits_{F\rightarrow\infty}\frac{T_F}{F}$, where $T_F$ is the duration of the fronthaul transmission and $T_E$ is the duration of the wireless transmission.
\end{definition}

\noindent Letting the fronthaul capacity $C_F$ scale as $r\log P$ with the power $P$, the parameter $r$ can be viewed as pre-log of the fronthaul capacity. We then define the NDT as follows.

\begin{definition}
(Normalized Delivery Time). For any achievable $\Delta(\mu,C_F,d,P)$, and given connectivity degree $d$, the NDT is defined as
\begin{equation}
\delta_{d}(\mu,r)=\lim\limits_{P\rightarrow\infty}\frac{\Delta(\mu,r\log{P},d,P)}{1/\log (P)}. \label{NDT}
\end{equation}
In addition, for any given pair of $(\mu,r)$ and a fixed $d$, the minimum NDT is defined as
\begin{equation}
\delta^*_{d}(\mu,r)=\inf\{\delta_{d}(\mu,r):\delta_{d}(\mu,r)\text{ is achievable}\}.
\end{equation}
Furthermore, we denote as $\delta_{E,d}(\mu,r)=\lim\limits_{P\rightarrow\infty}\frac{\Delta_E(\mu,r\log{P},d,P)}{1/\log (P)}$ and similarly as $\delta_{F,d}(\mu,r)=\lim\limits_{P\rightarrow\infty}\frac{\Delta_F(\mu,r\log{P},d,P)}{1/\log (P)}$ the achievable NDTs of wireless and fronthaul transmissions, respectively. The corresponding values achieved by an optimal scheme are defined as $\delta^*_{E,d}(\mu,r)$ and $\delta^*_{F,d}(\mu,r)$.
\end{definition}

As for the definition above, the NDT compares the delivery time per bit of the scheme of interest to that of an ideal interference-free system, which has the delivery time per bit of $1/\log(P)$ \cite{SenTanSim2017}. Hence, the NDT satisfies the inequality $\delta_d(\mu,r)\geq 1$. The minimum NDT $\delta^*_d(\mu,r)$ is convex in $\mu$ for any fixed value of $C_F$, as it can be proved by means of file-splitting and cache-sharing \cite[Lemma $1$]{SenTanSim2017}.

\textbf{Notation:} Throughout the paper, $[K]\triangleq\{1,\dots,K\}$ and $[N]\triangleq\{1,\dots,N\}$. $i,j,k$ and $n$ represent the index of the $i^{th},j^{th},k^{th}$ EN/user and the $n^{th}$ file. $W_n$ represents the $n^{th}$ message and $W_{[N]\backslash n}$ represents all messages in the library but the $n^{th}$ message.

\section{Cache-aided Blind Interference Avoidance \label{main}}
In this section, we present the proposed achievable NDT for the F-RAN system under study in the absence of fronthaul connections.
\begin{proposition} (Upper bound on minimum NDT)
For a $(K,d)$ regular network with no fronthauling and minimum storage, i.e., with $r=0$ and $\mu=1/d$, an upper bound on the minimum NDT is given as
\begin{equation}
\delta^*_{E,d}\left(\mu=\frac{1}{d},0\right)\leq \delta^{ach}_{E,d}=\begin{cases} \frac{2(d+1)\lceil\frac{d}{2}\rceil}{d}, & d\geq 2\\
1, & d=1. \end{cases} \label{NDTmine}
\end{equation}
\label{myresult}
\end{proposition}

The NDT described in Proposition \ref{myresult} is achieved by means of an MDS coded caching scheme and a carefully designed delivery scheme. For the caching scheme, each file is split into $d$ subfiles, and a $(K,d)$ MDS code is applied to create $K$ coded subfiles. Each EN stores one coded subfile for each file. This is done so that any user needs to obtain one subfile from each of its $d$ connected ENs, to recover its requested file. The delivery scheme is based on a novel blind interference avoidance scheme. The details of the proposed scheme are described in Appendix \ref{proofsection}, along with the proof of Proposition \ref{myresult}.

As a general remark, the role of intra-file coded caching in obtaining the NDT in (\ref{NDTmine}) should be emphasized. In fact, as discussed in \cite{SenTanSim2017}, in the presence of full connectivity and CSI, intra-file coded caching can only bring minor improvements in the NDT, which are limited to a factor of $2$ for any F-RAN parameters. In contrast, with partial connectivity and no CSI, coded caching is instrumental in reducing the delivery latency. In fact, it can be seen that with $\mu=1/d$, it is not possible to obtain a finite NDT with uncoded caching as long as the inequality $K>d$ holds, even with full CSI.

As a benchmark, we now compare the NDT of the proposed scheme to the NDT in \cite{YiGes2015}. This is the best known scheme for a $(K,d)$ regular network with no CSI \cite[Theorem $4$]{YiGes2015}. For a $(K,d)$ regular network with full storage, i.e., with $\mu=1$, the scheme achieves the NDT
\begin{equation}
\delta^{full}_{E,d}=\begin{cases} 
\frac{d+1}{2}, & d\geq 2\\
1, & d=1.
\end{cases} \label{NDTG}
\end{equation}

\noindent The achievable scheme in \cite{YiGes2015} uses blind interference alignment with full EN cooperation. In order to achieve the NDT in (\ref{NDTG}), the coherence time of the channel must be greater than $d+1$. Under such an assumption, the ENs break each desired file into two uncoded subfiles, and design precoding vectors jointly so that the interference only affects $d-1$ dimensions out of $d+1$ dimensions at each user. The remaining two dimensions are then reserved for the two uncoded desired subfiles.

While the scheme in \cite{YiGes2015} assumes full caching, whereby each EN stores all files, we note here that, when using intra-file coded caching, the NDT in (\ref{NDTG}) can be achieved with a smaller cache storage. Specifically, consider a $(K,2)$ MDS coded caching scheme, where each file is split into two uncoded subfiles, so that $K$ coded subfiles are created. Instead of storing the entire file, each EN stores only one coded subfile for each file. This reduces the required cache size to $\mu=1/2$, while still ensuring that the scheme in \cite{YiGes2015} can be implemented. Therefore, the scheme in \cite{YiGes2015} can, in fact, be applied for any value of $\mu$ between $1/2$ and $1$.

\begin{corollary}
The proposed scheme achieves an NDT that is within a multiplicative factor of $4$ of the best known scheme with full caching \cite{YiGes2015}, i.e.,
\begin{equation}
\frac{\delta^{ach}_{E,d}}{\delta^{full}_{E,d}}=\frac{4\lceil\frac{d}{2}\rceil}{d}\leq \frac{4(\frac{d}{2}+1)}{d}= 2+\frac{4}{d}\stackrel{(a)}{\leq} 4, \label{ratio}
\end{equation}
where inequality $(a)$ holds for $d\geq 2$.
\end{corollary}

We next demonstrate the proposed scheme through an example.

\begin{figure}[t]
\centering
	\includegraphics[width=0.65\linewidth]{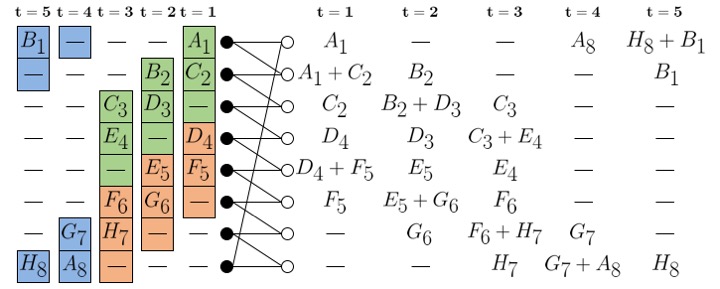}
	\caption{Illustration of the proposed scheme for a $(8,2)$ regular network.
	\label{example82}}
	\vspace{-20pt}
\end{figure}

\begin{example} \label{example82e}
Let us consider an $(8,2)$ regular network with $\mu=1/2$ (see Fig. \ref{example82}). For the caching scheme, each file is split into two subfiles, so that eight coded subfiles are created. We now describe the proposed blind interference avoidance scheme for this example. Note that each user can decode the desired file with any two coded subfiles. The goal is to deliver two coded subfiles to every user using as few time slots as possible. Denote the requested file vector as $\overline{D}=(A,B,C,D,E,F,G,H)$ and with subscript $i$ the index of the coded subfile, so that $\en_i$ has available $(A_i,B_i,C_i,D_i,E_i,F_i,G_i,H_i)$. With reference to Fig. \ref{example82} for an illustration, for $t=1$, $\en_1$ transmits $A_1$ to $\uu_1$. File $A_1$ is seen at $\uu_2$ as interference. In order to maximize the amount of users being served in the same time slot, $\en_2$ transmits $C_2$ to $\uu_3$. For $\uu_3$ to decode $C_2$, $\en_3$ needs to be silent to avoid interference. Similarly, $\en_4$ sends $D_4$ to $\uu_4$, and $\en_5$ sends $F_5$ to $\uu_6$ while $\en_6$ stays silent. For $t=2$, we shift the scheduled ENs downward by one EN serving $\uu_2,\uu_4,\uu_5$ and $\uu_7$. The ENs are scheduled in a similar way for $t=3$ serving $\uu_3,\uu_5,\uu_6$ and $\uu_8$.

However, for $t=4$, $\en_4$ and $\en_5$ do not have new subfiles for $\uu_4$ and $\uu_6$. Therefore, they will stay silent, and only $\en_7$ and $\en_8$ are scheduled. Similarly, for $t=5$, only $\en_8$ and $\en_1$ are scheduled. A total of $16$ subfiles were delivered in $5$ time slots, yielding a sum DoF of $16/5$ and an NDT of $5/2$.
\end{example}

\section{Cache and Cloud-Aided Topological Interference Management \label{discussion}}
In this section, we present an achievable NDT in the presence of both fronthaul connections and EN caches. We consider either the proposed scheme that achieves $\delta^{ach}_{E,d}$ in Proposition \ref{myresult} or the scheme in \cite{YiGes2015} achieving the NDT $\delta^{full}_{E,d}$ in (\ref{NDTG}).

The proposed scheme requires a fractional cache capacity $\mu\geq 1/d$, for values $\mu<1/d$, we let the cloud send the remaining file fraction $(1/d-\mu)$ to the ENs in order to enable the proposed scheme. This yields the achievable NDT
\begin{equation}
\delta^{ach}_d(\mu,r)=d\times\frac{\frac{1}{d}-\mu}{r}+\delta^{ach}_{E,d},
\end{equation}
where the first term represents the fronthaul NDT, since each EN must receive a fraction $(1/d-\mu)$ for $d$ files.

As for the scheme in \cite{YiGes2015}, which requires $\mu\geq 1/2$, we similarly obtain for $\mu< 1/2$
\begin{equation}
\delta^{full}_d(\mu,r)=2\times\frac{\frac{1}{2}-\mu}{r}+\delta^{full}_{E,d},\label{NDTGwithcloud}
\end{equation}
since each EN must receive a fraction $(1/2-\mu)$ for two files from the cloud.

Note that, for values $1/d\leq\mu<1/2$, the cache size is large enough to implement the proposed scheme without the aid of the cloud. Hence, the achievable NDT is simply $\delta^{ach}_d(\mu,r)=\delta^{ach}_{E,d}$. To apply the scheme in \cite{YiGes2015}, we still need the cloud to send the remaining file fraction to the ENs, and the achievable NDT $\delta^{full}_d$ is given by (\ref{NDTGwithcloud}).

Furthermore, for values $\mu\geq 1/2$, both schemes can be implemented without the aid of fronthaul transmission. The achievable NDTs are $\delta^{ach}_d(\mu,r)=\delta^{ach}_{E,d}$ and $\delta^{full}_d(\mu,r)=\delta^{full}_{E,d}$, respectively. The main result of this section is the following.
\begin{proposition} For an F-RAN with $(K,d)$ regular edge topology, we have $\delta^{full}_d(\mu,r)\geq\delta^{ach}_d(\mu,r)$  when
\begin{align}
r\leq r_1=\frac{2d(d-2)\mu}{(d+1)(4\lceil\frac{d}{2}\rceil-d)}, &\quad and \quad 0<\mu<\frac{1}{d},\\
or \quad r\leq r_2=\frac{2d(1-2\mu)}{(d+1)(4\lceil\frac{d}{2}\rceil-d)}, &\quad and \quad \frac{1}{d}\leq\mu<\frac{1}{2}.
\end{align}
\label{pro2}
\end{proposition}

This result indicates that the proposed scheme is particularly useful when the availability of fronthaul capacity is limited. Intuitively, this is the case since the proposed scheme requires smaller values of the cache capacity $\mu$.

The proof of Proposition \ref{pro2} can be found in Appendix \ref{sec:AppenProp2}.

\section{Conclusions \label{conclusion}}
In this paper, we focused on cache and cloud enabled regular networks with partial wireless connections and no CSI. The main contribution of this work is the proposal of a novel cache-aided blind interference avoidance scheme. We showed that through intra-file coded caching, it is possible to obtain a finite NDT when the edge network is partially connected, with no CSI and minimum required cache size at the ENs. We further showed that the proposed scheme outperforms the state of the art when the network resources are limited, i.e., with low fronthaul capacity and/or low cache size. There are several interesting directions for future work. A first direction would be to obtain lower bounds on the minimum NDT and subsequently characterizing the optimal NDT in the presence of partial connectivity and no CSI. A second interesting direction would be to generalize the ideas presented herein for arbitrary (irregular) network topologies.

\appendix
\label{sec:proofs}
\subsection{Proof of Proposition \ref{myresult}} \label{proofsection}

In order to prove the achievability of the proposed scheme, we first present the MDS coded caching scheme, followed by the interference avoidance-based delivery scheme.

For the $(K,d)$ MDS coded caching strategy, each file $W_n$ is split into $d$ subfiles, i.e., $W_n=(W^{(1)}_n,W^{(2)}_n,\dots,W^{(d)}_n)$, where $W^{(i)}_n$ indicates the $i^{th}$ subfile of file $W_n$. We create $K$ coded subfiles out of these $d$ subfiles. Each coded subfile is of size $F/d$ bits, and each EN stores one coded subfile per file to meet the storage constraint $NF/d$ bits. Note that each UE can recover the desired file from any $d$ coded subfiles. We exploit this fact in the delivery scheme described next.

\begin{figure}[t]
\centering
  \begin{minipage}{0.35\textwidth}
  \vspace{-15pt}
    \includegraphics[width=\textwidth]{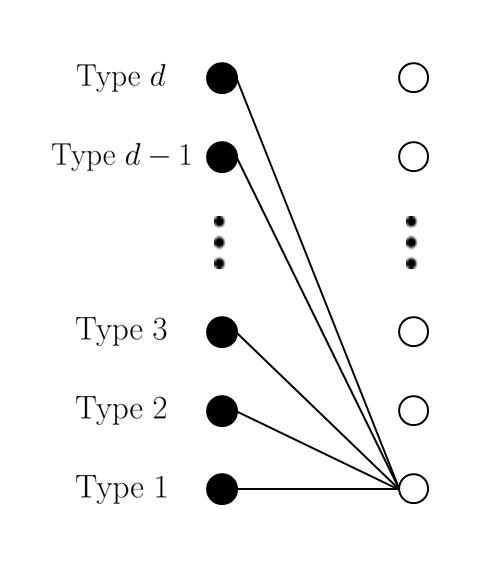}
    \vspace{-30pt}
    \caption{Subfile types.
    \label{type}}
  \end{minipage}
\hspace{1cm}
  \begin{minipage}{0.35\textwidth}
    \includegraphics[width=\textwidth]{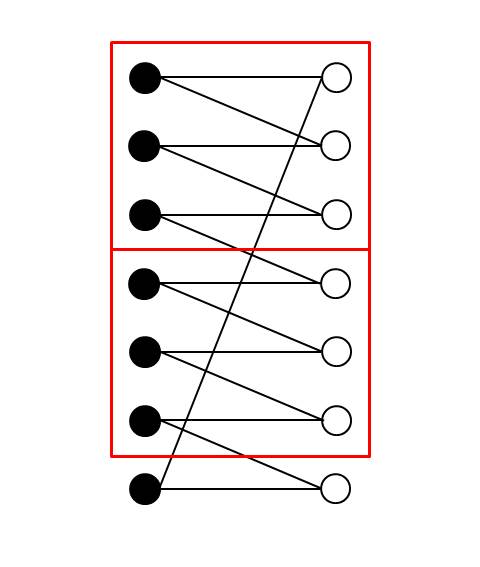}
    \vspace{-30pt}
    \caption{A $(7,2)$ regular network with two blocks.
    \label{block}}
  \end{minipage}
\vspace{-20pt}
\end{figure}

For delivery, we propose an interference avoidance-based strategy. Note that each user is connected to $d$ ENs, and each one of the $d$ ENs has an unique coded subfile for that particular user. We define these unique subfiles as types.

\begin{definition}
(Subfile Type). For any $(K,d)$ regular network with the proposed caching scheme, there are $d$ types of possible subfiles $\tau\in\{1,2,\dots,d\}$. We define the subfile for any $\uu_k$ from $\mbox{EN}_k$ as type $1$, the subfile for any $\uu_k$ from $\mbox{EN}_{k-1}$ as type $2$, and so on. Consequently, the subfile for $\uu_k$ from $\mbox{EN}_{k-d+1}$ is type $d$. Each type represents a distinct coded subfile. 
\end{definition}

We then exploit the topology considered in this paper to simultaneously schedule as many ENs and users as possible, and eventually ensuring that all subfile types are received at all users. We briefly illustrate the notion of subfile type and blind interference avoidance through an example.

\begin{example}
For a $(7,2)$ regular network (see Fig. \ref{block}), each file is split into two subfiles, so that seven coded subfiles are created. Each EN stores one distinct coded subfile per file. To avoid interference, $\mbox{EN}_{1}$ is scheduled to send type $1$ subfile to $\uu_1$, and $\mbox{EN}_{2}$ is scheduled to send type $2$ subfile to $\uu_3$ while $\mbox{EN}_{3}$ is scheduled to be silent. Similarly, $\mbox{EN}_{4}$ is scheduled to send type $1$ subfile to $\uu_4$, and $\mbox{EN}_{5}$ is scheduled to send type $2$ subfile to $\uu_6$ while $\mbox{EN}_{6}$ is scheduled to be silent. $\mbox{EN}_{7}$ has to be silent to avoid interference at $\uu_1$ (complete examples will be shown later in this section).
\end{example}

Clearly, in this example, three consecutive ENs need to be scheduled jointly to maximize the amount of users being served simultaneously. We denote these ENs as forming a block as defined next.

\begin{definition}
(Block). A block refers to a group of ENs and users that are scheduled together to avoid interference at the intended users. Each block consists of $d+1$ EN and user pairs. There are at most $\lfloor\frac{K}{d+1}\rfloor$ blocks.
\end{definition}

As shown in the previous example, two subfile types are delivered within each block to two different users. We define a series of transmissions that ensures every user to get those two subfile types as a stage.

\begin{definition}
(Stage) A series of transmissions at each Stage $s$, where $s\in\{1,2,...,\lceil\frac{d}{2}\rceil\}$, focuses on sending subfile types $s$ and $d-s+1$ to all users.
\end{definition}

\begin{figure}[t]
\centering
	\includegraphics[width=0.65\linewidth]{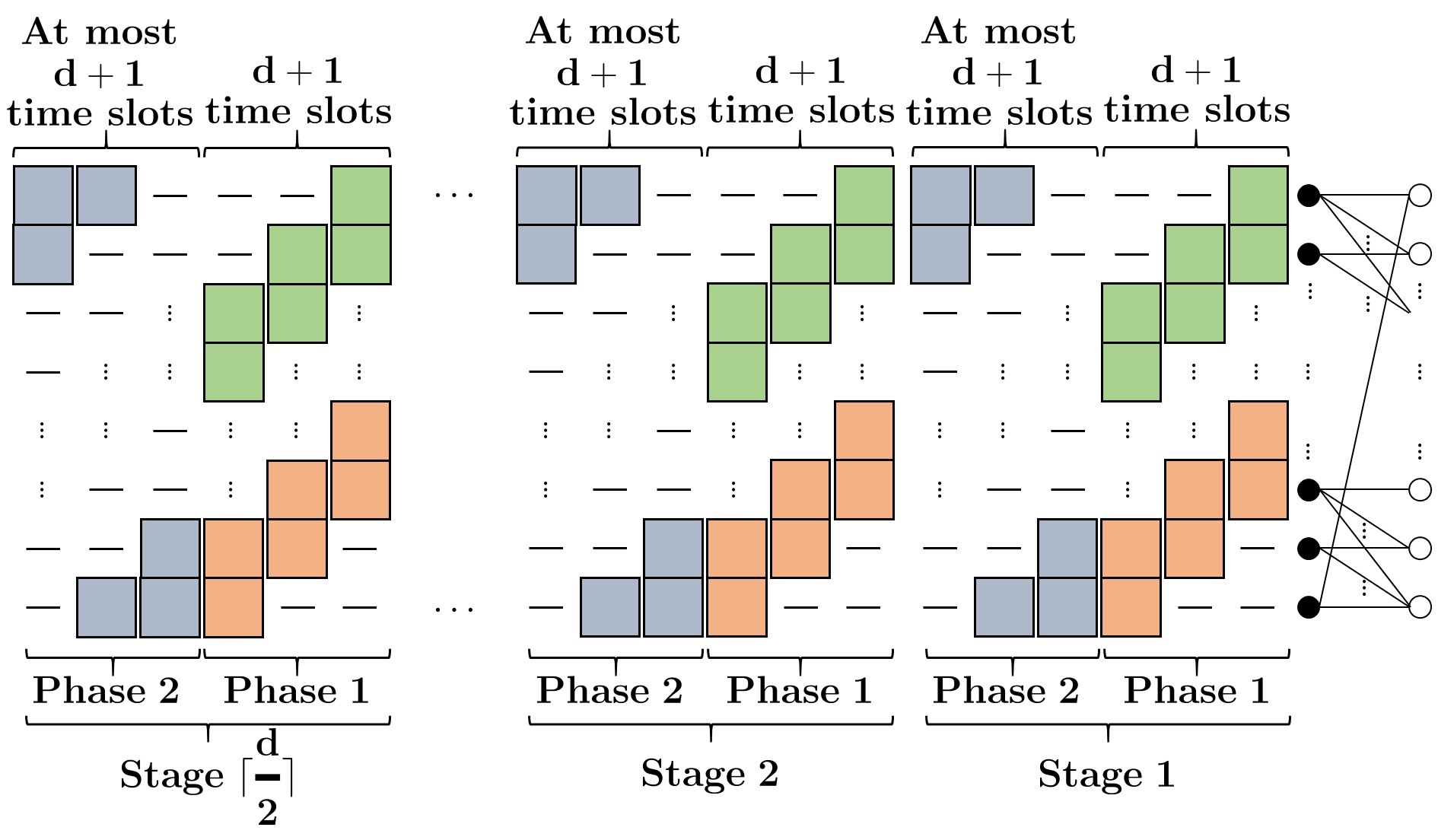}
	\caption{A general view of stages and phases, and their respective durations.
	\label{stage}}
	\vspace{-20pt}
\end{figure}

To deliver type $s$ and type $d-s+1$ subfiles to all the user, we can shift blocks in a cyclic manner. However, if we simply shift the blocks $d+1$ number of times, we will eventually arrive at a point where ENs have no new subfiles for most of users and send redundant subfiles. Thus, we break each stage into two phases to avoid such a scenario (see Fig. \ref{stage}).

\noindent $(a)$ \textbf{Phase $1$:} At $t$-$th$ time slot of Phase $1$ of Stage $s$, we schedule $\mbox{EN}_i$ to deliver type $s$ subfile to user $\uu_{i+(s-1)}$, and schedule $\mbox{EN}_{i+s}$ to deliver type $d-s+1$ subfile to user $\uu_{i+d}$, where $i=\{t,t+(d+1),\dots,t+(\lfloor\frac{K}{d+1}\rfloor-1)(d+1)\}$. A total of $d+1$ time slots are needed for Phase $1$, i.e. $t\in\{1,\dots,(d+1)\}$, because once it reaches the $(d+2)$-$th$ time slot, most of ENs do not have new coded subfiles about the desired files for intended users.

\noindent $(b)$ \textbf{Phase $2$:} When $K$ is not divisible by $d+1$, at the end of Phase $1$ of any Stage $s$, $\uu_i$, where $i=\{d,d-1,\dots,d-(K\mod (d+1))+1\}$, will only receive type $s$ subfiles. Last $(K\mod (d+1))$ users will only receive type $d-s+1$ subfiles, i.e. $\uu_i$, where $i=\{K-(K\mod (d+1))+1,\dots,K\}$. Therefore, at $t$-$th$ time slot of Stage $s$, where $t>d+1$, we schedule $\mbox{EN}_i$ to deliver type $s$ subfiles to $\uu_{i+(s-1)}$, and schedule $\mbox{EN}_{i+s}$ to deliver type $d-s+1$ subfiles to user $\uu_{i+d}$, where $i=\{t+(\lfloor\frac{K}{d+1}\rfloor-1)(d+1)\}$. The duration of Phase $2$ is $(K\mod (d+1))$ time slots, which does not exceed $d+1$ time slots. 

Hence, any stage will take at most $2(d+1)$ time slots. The number of stages can be easily obtained by counting how many subfile types can be paired, i.e., $\lceil\frac{d}{2}\rceil$. In sum, $K\cdot d$ subfiles can be sent in $\lceil\frac{d}{2}\rceil$ stages, and each stage occupies at most $2(d+1)$ time slots. This gives a lower bound on the maximum sum DoF of $Kd/2(d+1)\lceil\frac{d}{2}\rceil$ and yield an upper bound on the minimum NDT of $2(d+1)\lceil\frac{d}{2}\rceil/d$. This completes the proof of Proposition \ref{myresult}.

One might notice that after Stage $2$, the range of users suffers from interference widen. Users will start seeing interference from the previous block. Lemma \ref{blocksize} shows that this effect does not cause interference at intended users.

\begin{lemma} \label{blocksize}
(Impact of interference on intended users). As long as the subfile types delivered at the same time sum up to $d+1$, the intended users will not suffer from the interference coming from the preceding block.
\end{lemma}

\begin{figure}[t]
\begin{center}
	\includegraphics[width=0.45\linewidth]{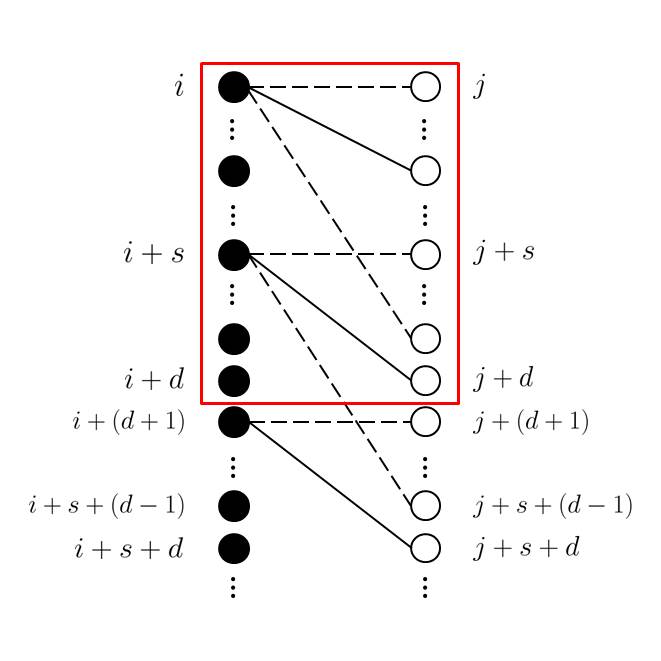}
	\vspace{-20pt}
	\caption{A block of ENs and users, and its interference to the next block. Solid lines represent desired link and dash lines represent interference, with indices on the side and $i=j$.
	\label{range}}
\end{center}
\vspace{-30pt}
\end{figure} 

We next present an example with two stages to show that Lemma \ref{blocksize} is indeed true and defer the proof of Lemma \ref{blocksize} to Appendix \ref{sec:AppenLemma1}.

\begin{figure}[t]
\centering
  \begin{minipage}{0.9\textwidth}
    \includegraphics[width=\textwidth]{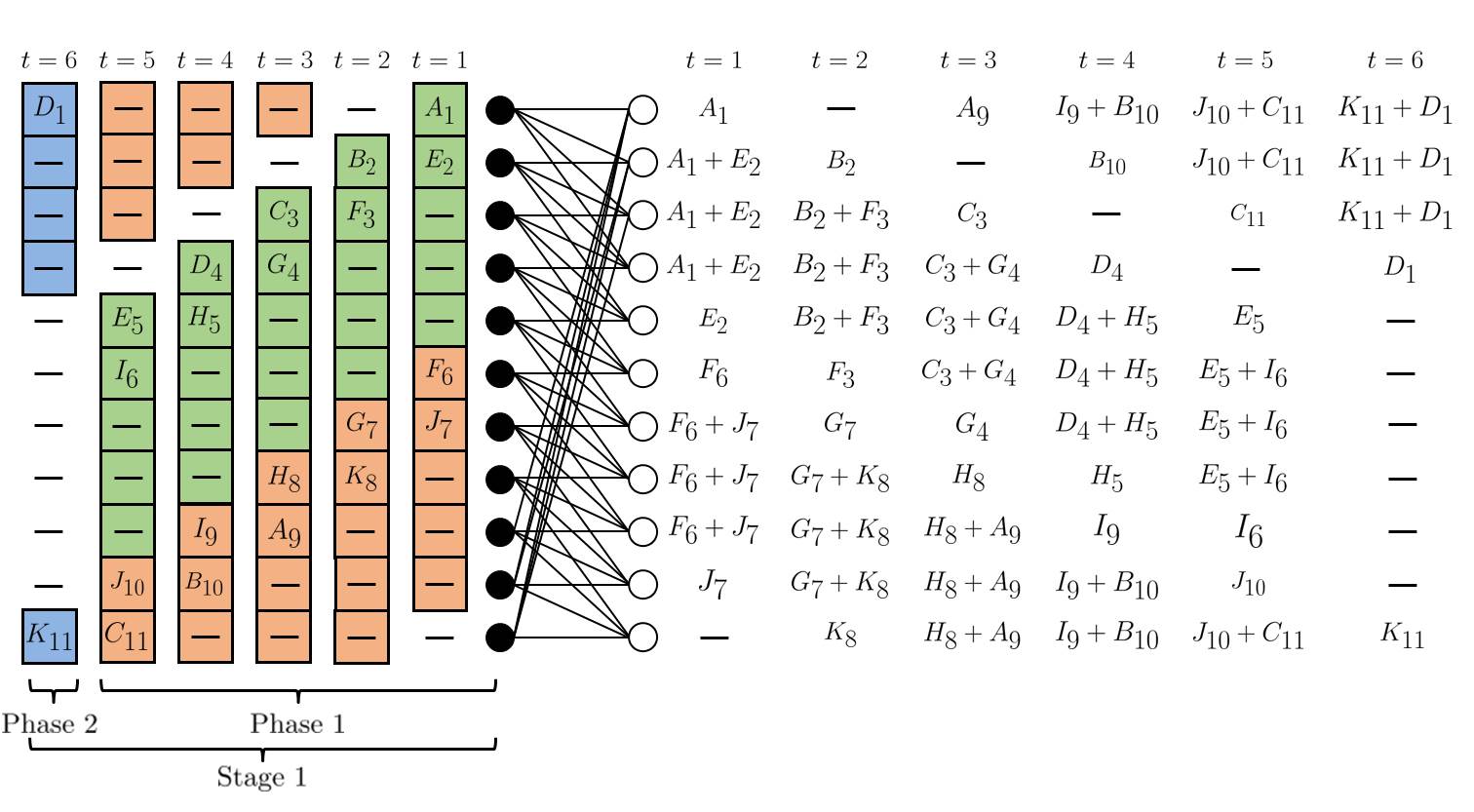}
    \caption{Edge transmission in Stage $1$ for a $(11,4)$ regular network with cache size $\mu=1/4$.\\
    \label{example1}}
  \end{minipage}
\hfill
  \begin{minipage}{0.9\textwidth}
    \includegraphics[width=\textwidth]{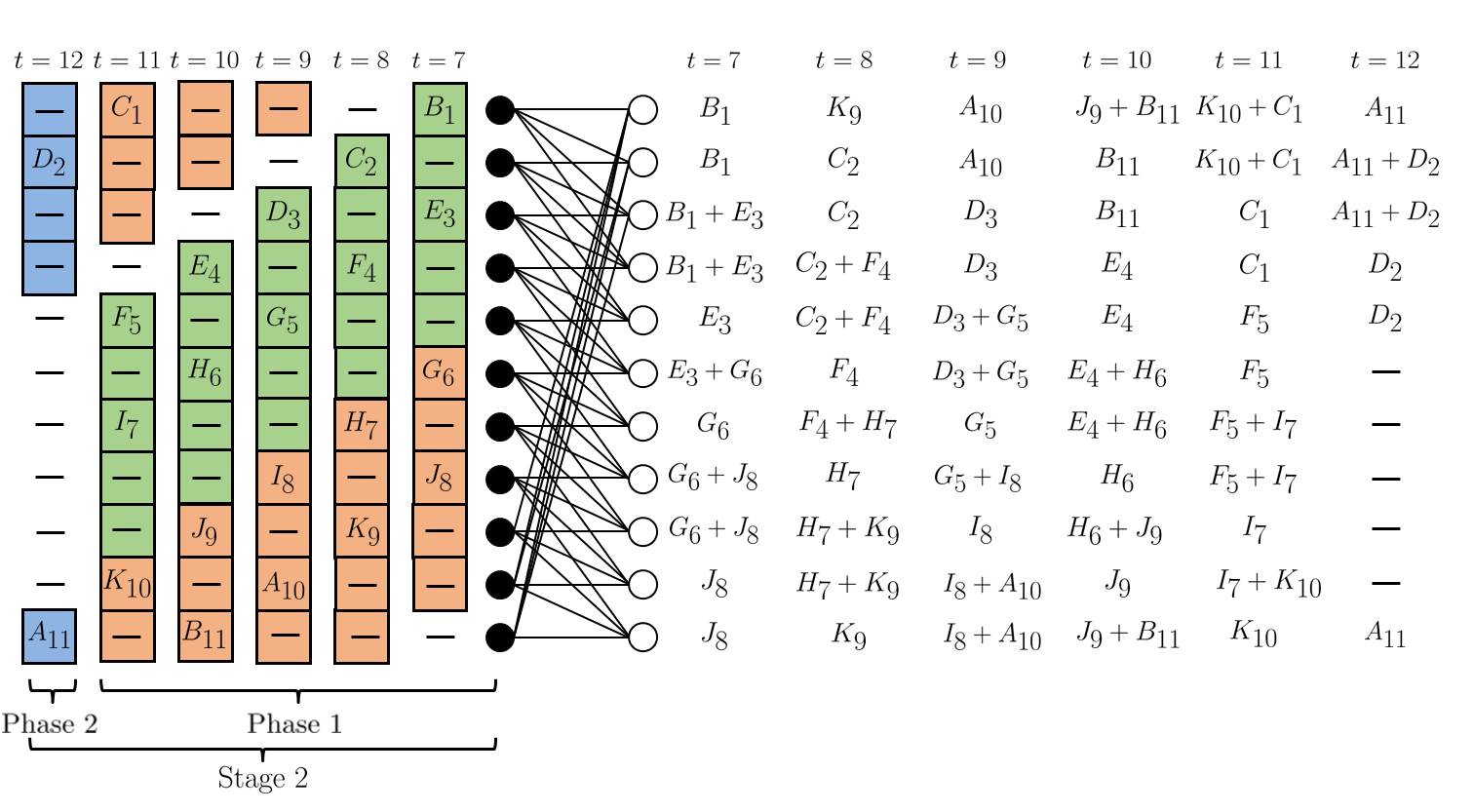}
    \caption{Edge transmission in Stage $2$ for a $(11,4)$ regular network with cache size $\mu=1/4$.
    \label{example2}}
  \end{minipage}
  \vspace{-20pt}
\end{figure}

\begin{example} \label{example}
For a $(11,4)$ regular network, there are two blocks of size $d+1=5$ and $2$ stages. At Stage $1$, type $1$ and type $4$ subfiles will be served to all users. During Stage $1$ Phase $1$ at $t=1$ (see Fig. \ref{example1}), $\mbox{EN}_1$ sends $A_1$ (type $1$ subfile) to $\uu_1$, $\mbox{EN}_2$ sends $E_2$ (type $4$ subfile) to $\uu_5$, $\mbox{EN}_6$ sends $F_6$ (type $1$ subfile) to $\uu_6$, and $\mbox{EN}_7$ sends $I_7$ (type $4$ subfile) to $\uu_{10}$. Notice that at the end of $t=1$, $\uu_1$ and $\uu_6$ received type $1$ subfiles, $\uu_5$ and $\uu_{10}$ received type $4$ subfiles. At $t=2$, $\mbox{EN}_2,\mbox{EN}_3,\mbox{EN}_7,\mbox{EN}_8$ transmit type $1$ and type $4$ subfiles to user $\uu_2,\uu_6,\uu_7,\uu_{11}$ in a similar fashion. Same form of transmissions are repeated for $5$ time slots in order to cover type $1$ and type $4$ subfiles for most of users. When $t=5$, $\uu_4$ has not received its type $4$ subfile and $\uu_{11}$ has not received its type $1$ subfile. Therefore, we enter Stage $1$ Phase $2$, where $\mbox{EN}_1,\mbox{EN}_{11}$ send $D_1$ and $K_{11}$ to $\uu_4$ and $\uu_{11}$, respectively. Phase $2$ takes $1$ time slot. Stage $1$ takes a total of $6$ time slots.

We then move onto Stage $2$ Phase $1$ (see Fig. \ref{example2}), in which we send type $2$ and type $3$ subfiles together. Note that we no longer schedule two adjacent ENs now. At $t=7$, $\mbox{EN}_1$ transmits $B_1$ (type $2$ subfile) to $\uu_2$, $\mbox{EN}_3$ transmits $E_3$ (type $3$ subfile) to $\uu_5$. Same form of transmissions is again repeated for $5$ time slots. In Phase $2$, $D_2$ and $A_{11}$ are sent by $\mbox{EN}_2,\mbox{EN}_{11}$ to $\uu_4$ and $\uu_1$, respectively. Therefore, we send a total of $44$ subfiles in $12$ time slots and achieve a sum DoF of $11/3$ and an NDT of $3$.
\end{example}

\subsection{Proof of Lemma~\ref{blocksize}} \label{sec:AppenLemma1}
Since every time slot is just a shifted version of the previous time slot and ENs are scheduled in the same way in each block, we focus on just one single time slot and any two consecutive blocks. During Phase $1$ of any Stage $s$, we send type $s$ and type $d-s+1$ subfiles, where the sum of them is $d+1$.
\begin{itemize}
\item ENs that will be scheduled are $\mbox{EN}_i,\mbox{EN}_{i+s},\mbox{EN}_{i+(d+1)}$ and $\mbox{EN}_{i+s+(d+1)}$. However, since we are interested in the impact of interference between any two consecutive blocks, we focus on $\mbox{EN}_{i+s}$ and $\mbox{EN}_{i+(d+1)}$ only.
\item The receiver sets of $\mbox{EN}_{i+s},\mbox{EN}_{i+(d+1)}$ are $\mathcal{R}_{i+s}=\{j+s,\dots,j+s+(d-1)\}$ and $\mathcal{R}_{i+(d+1)}=\{j+(d+1),\dots,j+2d+1\}$. Their respective intended users are $\uu_{j+d}$ and $\uu_{j+s+d}$.
\item As shown in Fig. \ref{range}, when $s\geq 2$, $\mathcal{R}_{i+s}$ overlaps with $\mathcal{R}_{i+(d+1)}$. Users who see signals from both ENs are $\mathcal{R}_{i+s}\cap\mathcal{R}_{i+(d+1)}=\{j+(d+1),\dots,j+s+(d-1)\}$. It can be seen that the intended user of $\mbox{EN}_{i+(d+1)}$, $\uu_{j+s+d}$, is not in the set.
\end{itemize}
Thus, we conclude that the intended user in the following block will not see interference.

\subsection{Proof of Proposition~\ref{pro2}}\label{sec:AppenProp2}
\begin{figure}[htb]
\centering
  \begin{minipage}{0.45\textwidth}
    \includegraphics[width=\textwidth]{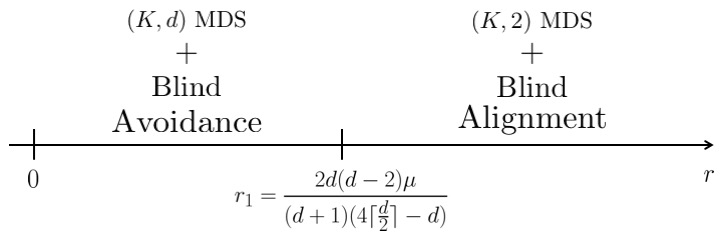}
  \end{minipage}
\hspace{1cm}
  \begin{minipage}{0.45\textwidth}
    \includegraphics[width=\textwidth]{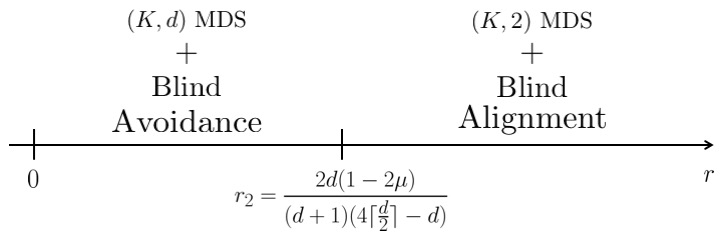}
  \end{minipage}
  \caption{Choice of caching and delivery schemes for any $(K,d)$ regular network. (Left) $\mu\in[0,1/d)$. (Right) $\mu\in[1/d,1/2)$.\\
    \label{rboundary1}}
    \vspace{-20pt}
\end{figure}

Recall that in Section \ref{discussion}, we have the following results. For values $\mu< 1/d$, the achievable NDTs of the proposed scheme and the scheme in \cite{YiGes2015}, both with fronthaul connections, are
\begin{equation}
\delta^{ach}_d(\mu,r)=d\times\frac{\frac{1}{d}-\mu}{r}+\delta^{ach}_{E,d},\quad \delta^{full}_d(\mu,r)=2\times\frac{\frac{1}{2}-\mu}{r}+\delta^{full}_{E,d},
\end{equation}
respectively.
To show that the proposed scheme is a better choice when the fronthaul capacity is low, we obtain a threshold $r_1$ on the fronthaul capacity by comparing $\delta^{ach}_d(\mu,r)$ to $\delta^{full}_d(\mu,r)$ as follows.
\begin{align}
\delta^{ach}_d(\mu,r)&\leq\delta^{full}_d(\mu,r)\\
d\times\frac{\frac{1}{d}-\mu}{r} + \frac{2(d+1)\lceil\frac{d}{2}\rceil}{d} &\leq 2\times\frac{\frac{1}{2}-\mu}{r} + \frac{d+1}{2}\\
\frac{2(d+1)\lceil\frac{d}{2}\rceil}{d} - \frac{d+1}{2} &\leq \frac{1-2\mu}{r} - \frac{1-d\mu}{r}\\
\frac{(d+1)(4\lceil\frac{d}{2}\rceil -d)}{2d} &\leq \frac{(d-2)\mu}{r}\\
r &\leq \frac{2d(d-2)\mu}{(d+1)(4\lceil\frac{d}{2}\rceil-d)}=r_1. \label{r1}
\end{align}
We conclude that the proposed scheme yields a lower NDT when the fronthaul capacity is lower or equal to $r_1$ for $\mu<1/d$.

Similarly, for values $1/d\leq\mu<1/2$, the achievable NDTs of the propsed scheme and the scheme in \cite{YiGes2015} are
\begin{equation}
\delta^{ach}_d(\mu,r)=\delta^{ach}_{E,d},\quad \delta^{full}_d(\mu,r)=2\times\frac{\frac{1}{2}-\mu}{r}+\delta^{full}_{E,d},
\end{equation}
respectively. We obtain a threshold $r_2$ on the fronthaul capacity for this cache size regime by comparing $\delta^{ach}_d(\mu,r)$ to $\delta^{full}_d(\mu,r)$.
\begin{align}
\delta^{ach}_d(\mu,r)&\leq\delta^{full}_d(\mu,r)\\
\frac{2(d+1)\lceil\frac{d}{2}\rceil}{d} &\leq 2\times\frac{\frac{1}{2}-\mu}{r} + \frac{d+1}{2}\\
\frac{2(d+1)\lceil\frac{d}{2}\rceil}{d} - \frac{d+1}{2} &\leq \frac{1-2\mu}{r}\\
\frac{(d+1)(4\lceil\frac{d}{2}\rceil -d)}{2d} &\leq \frac{1-2\mu}{r}\\
r &\leq \frac{2d(1-2\mu)}{(d+1)(4\lceil\frac{d}{2}\rceil-d)}=r_2. \label{r2}
\end{align}
It can be seen that a lower NDT can be achieved with the proposed scheme when $r\leq r_2$ for $1/d\leq\mu<1/2$.

For values $\mu\geq 1/2$, both schemes can be implemented without requesting any fraction of the files from the cloud. Thus, the achievable NDTs are
\begin{equation}
\delta^{ach}_d(\mu,r)=\delta^{ach}_{E,d},\quad \delta^{full}_d(\mu,r)=\delta^{full}_{E,d},
\end{equation}
respectively. To this end, the scheme in \cite{YiGes2015} achieving the NDT $\delta^{full}_{E,d}$ will always yield a lower NDT than the proposed scheme. Thus, we do not have a threshold on the fronthaul capacity for this cache size regime. This completes the proof of the achievability when cloud is enabled.

\bibliographystyle{IEEEtran}
\bibliography{Ref}

\end{document}